**Title:** Interferometric 4D-STEM for Lattice Distortion and Interlayer Spacing Measurements in Bilayer and Trilayer Two-dimensional Materials

*Michael J. Zachman\*, Jacob Madsen, Xiang Zhang, Pulickel M. Ajayan, Toma Susi, Miaofang Chi\**

Dr. M. J. Zachman, Dr. M. Chi
Center for Nanophase Materials Sciences, Oak Ridge National Laboratory, Oak Ridge, TN, 37831, USA
E-mail: zachmanmj@ornl.gov, chim@ornl.gov

Dr. J. Madsen, Prof. T. Susi
Faculty of Physics, University of Vienna, Boltzmanngasse 5, Vienna, 1090, Austria

Dr. X. Zhang, Prof. P. M. Ajayan
Department of Materials Science and NanoEngineering, Rice University, Houston, TX, 77005, USA

Van der Waals materials composed of stacks of individual atomic layers have attracted considerable attention due to their exotic electronic properties that can be altered by, for example, manipulating the twist angle of bilayer materials or the stacking sequence of trilayer materials. To fully understand and control the unique properties of these few-layer materials, a technique that can provide information about their local in-plane structural deformations, twist direction, and out-of-plane structure is needed. In principle, interference in overlap regions of Bragg disks originating from separate layers of a material encodes three-dimensional information about the relative positions of atoms in the corresponding layers. Here, we describe an interferometric four-dimensional scanning transmission electron microscopy technique that utilizes this phenomenon to extract precise structural information from few-layer materials with nm-scale resolution. We demonstrate how this technique enables measurement of local pm-scale in-plane lattice distortions as well as twist direction and average interlayer spacings in bilayer and trilayer graphene, and therefore provides a means to better understand the interplay between electronic properties and precise structural arrangements of few-layer 2D materials.

## 1. Introduction

The electronic properties of two-dimensional (2D) materials are highly dependent on the precise structure of the material, both in-plane within a single layer and out-of-plane between layers in stacked 2D materials.[1–18] A noteworthy example is twisted bilayer graphene, which can not only transition to a superconducting state at a specific "magic" twist angle,[19] but has also been shown both theoretically and experimentally to undergo an in-plane structural reconstruction at low twist angles, significantly altering the electronic state of the material from that of rigidly twisted material.[16] In addition, the stacking sequence and relative orientation of layers in trilayer graphene have also recently been shown to enable the electronic character of the material to be tuned, producing a semimetal in the ABA configuration and a semiconductor in the ABC configuration, for example.[16] Since the often exotic electronic properties of these materials are so exquisitely dependent on their structural configuration, the ability to precisely determine this configuration is therefore key to fully understanding and controlling their properties.

Several techniques have been described that enable various structural properties of bilayer materials to be measured in both supported and freestanding layers, for example, by scanning tunneling microscopy (STM),[20] dark-field transmission electron microscopy (DF-TEM),[21] and scanning transmission electron microscopy (STEM).[22–24] While these techniques generate a large amount of valuable structural information, each has their own set of strengths and weaknesses. For example, conventional annular dark-field (ADF) STEM imaging can provide high-resolution structural information, but low signal strength for light elements such as carbon can make precise positional fitting of individual atoms challenging in materials like graphene. In addition, conventional TEM and STEM generally cannot distinguish which layer an imaged atom resides within in a bilayer material, for example, making measurement of structural distortions through atom position tracking challenging.

A large effort has recently been put forth to address some of these limitations. For example, an innovative TEM diffraction technique was recently demonstrated for determining average deformations due to structural reconstructions in bilayer graphene at low twist angles.[22,25–28] While this technique is able to provide average lattice deformation information, determining the sample structure involves comparisons between experimental and simulated diffraction patterns generated from model structures. In addition, this technique is limited to maximum twist angles of between 2° and 3°, and similar to other projection techniques, it does not provide information about out-of-plane structure. Very recently, a four-dimensional STEM (4D-STEM) technique based on Bragg interferometry was demonstrated that allowed in-plane structure distortions to be measured in bilayer graphene.[29] While this provides the valuable ability to resolve local structural distortions, it was applied only to low-twist angle materials, up to 1.6°, and again cannot provide information about out-of-plane structure, such as the interlayer spacings. Finally, none of these projection techniques can provide information about the direction of twist of one layer with respect to the other, which could be useful for studying chiral materials.[30,31] Therefore, to fully understand the precise structural arrangements of few-layered materials, a technique is needed that can directly and precisely reveal pm-scale in-plane structural information across a range of twist angles with high spatial resolution, as well as provide information about the twist direction and out-of-plane structure.

Here, we describe and provide an initial demonstration of an interferometric four-dimensional STEM (4D-STEM) technique that utilizes a defocused probe and interference between overlapping Bragg disks that originate from separate layers of a few-layered material to directly provide information about the relative positions of atoms in the separate layers comprising few-layer 2D materials, such as bilayer and trilayer graphene. In principle, this technique provides access to the full three-dimensional structure of the sample. In addition to demonstrating how this enables pm-scale in-plane structural deformations to be mapped with nm-scale resolution, we show how it permits the twist direction to be measured, as well as the average interlayer spacing, which allows the stacking sequence of trilayer materials to be directly determined in certain cases.

## 2. Results and Discussion

STEM techniques typically generate data in projection, compressing information about a 3D material into data in 2D. As a consequence, disentangling atom positions from different layers of a stacked 2D material, and therefore understanding their relative positions, is not straightforward using conventional STEM techniques. Here, we describe a nanodiffraction-based 4D-STEM technique that utilizes interference between overlapping Bragg disks that originate from separate layers of a stacked 2D material to reveal the relative positions of the atoms within the layers. Such Bragg interference has been described previously and was utilized to reconstruct, for example, stretching and corrugation of a hexagonal boron nitride-graphene heterostructure over large fields of view by convergent beam electron holography.[16,32,33] Our approach is to pair this Bragg interference with the inherent spatial resolution of STEM to enable the relative position of atoms in separate layers to be directly mapped at the nanometer scale by 4D-STEM.

A STEM probe focused on a sample produces flat Bragg disks at fixed scattering angles, as shown in the left panel of **Figure 1a**. For bilayer graphene with a twist angle near 3°, this produces a diffraction pattern such as the one simulated in Figure 1b. If the sample is positioned away from the focal plane of the probe (or a defocused probe is used), each of the diffracted beams focuses in the same plane as the transmitted beam but displaced in real space in the direction of the corresponding Bragg disks. For an underfocused beam, the focal points of the diffracted beams are real and for an overfocused beam they are virtual, as shown in the middle and right panels of Figure 1a, respectively. The distance between the focal points of the scattered and transmitted beams, denoted $D$ in Figure 1a, increases with increasing defocus. For a monolayer sample, small defocus values produce a negligible effect on the Bragg disks since the diffraction pattern is observed in the far-field (and hence only diffraction angle matters). For a bilayer sample, however, each layer of the material diffracts separately, allowing for interference between the Bragg disks originating from the two layers. For example, if a twist angle exists between the layers, such as in Figure 1, the direction of diffraction from the layers differs by this angle. For small enough angles or large enough Bragg disks, an overlap of the disks originating from the individual layers takes place, as in Figure 1b. When the probe is focused out of the sample plane, a small difference in the position of the diffracted beam focal points is produced that is approximately perpendicular to the direction of the transmitted beam focal point, as shown in Figure 1c. These spatially distinct focal points then act as coherent point sources that propagate wavefronts to the diffraction plane. This results in interference in the diffraction plane, visible where the Bragg disks corresponding to the point sources overlap, as shown schematically in Figure 1d and in the full simulated diffraction pattern in Figure 1e. In principle, these interference fringes contain all the information necessary to determine the local in-plane structural arrangement and twist angle, as well as interlayer spacing through the phase, wavelength, and angle of the fringes. In this work, we demonstrate how the phase of the fringes can be used to probe the local in-plane structural arrangement and the angle can be used to determine the twist direction and average interlayer separation of the material.

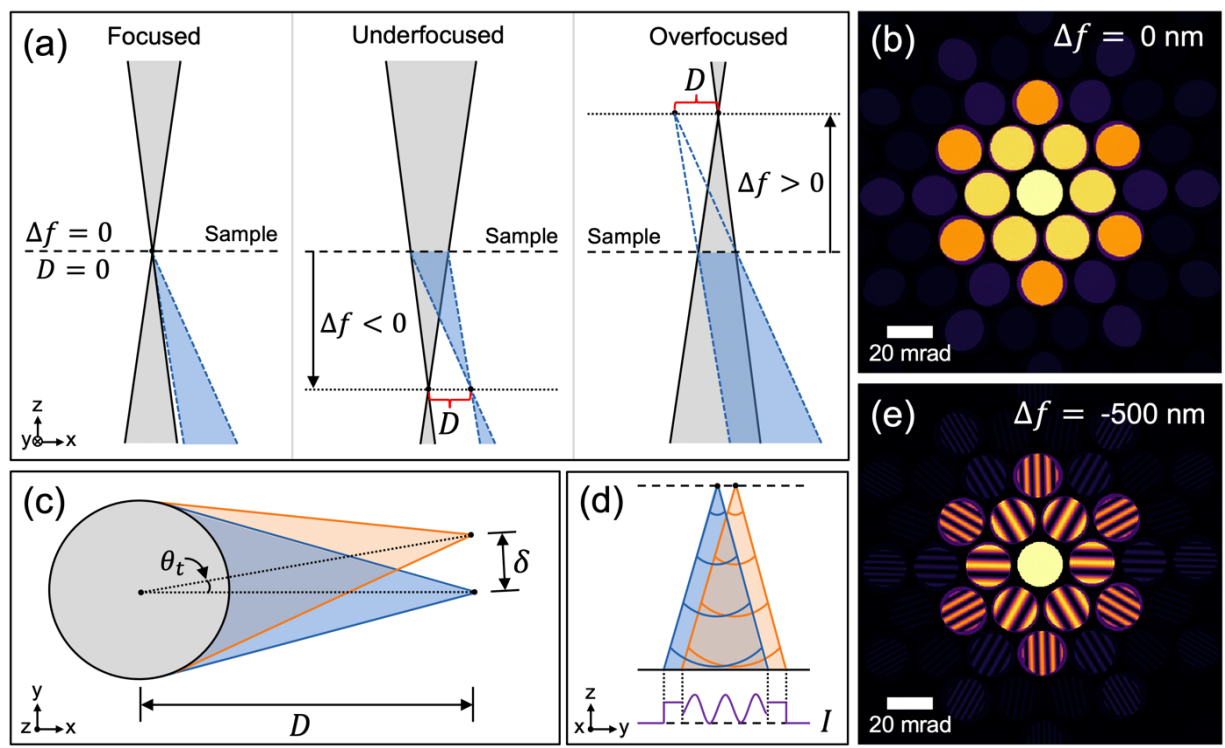

**Figure 1.** Schematics and simulated diffraction patterns depicting the origin and appearance of interference fringes in overlapping regions of Bragg disks that originate from separate layers of twisted bilayer graphene. (a) Schematics for diffraction from an in-focus, underfocused, and overfocused STEM probe, depicted in the left, center, and right panels, respectively. (b) An in-focus probe results in nearly flat Bragg disks in the diffraction plane. (c) Twisted bilayer materials result in separate real-space locations for focal points of diffracted beams originating in separate layers. (d) These separate focal points act as coherent point sources that can interfere in the diffraction plane where their corresponding Bragg disks overlap. (e) A defocused probe results in interference fringes with a shorter wavelength, visible within the Bragg disk, that overlap for a large enough defocus/convergence angle.

The relationship between the phase of the Bragg interference fringes and the sample structure can be understood by utilizing the concept of the structure factor, which relates the amplitude and phase of the diffracted beams to the type and positions of atoms in the unit cell.[34] For thin samples where kinematical scattering dominates, diffraction from a material can be defined as a lattice sum, which is a function of the repeating unit cell and defines the Bragg disk positions, multiplied by a structure factor that can be defined as follows:

$$F = \sum_j e^{-i\vec{k}\cdot\vec{r_j}} f_j(\vec{k}),\tag{1}$$

where $\vec{k}$ is the scattering vector, $\vec{r}$ is the real-space position of an atom in the unit cell, $f_j$ is the atomic scattering factor of that atom, and the index, $j$, refers to atoms in the unit cell. Through this relationship, the relative phases of the diffracted Bragg disks are tied to the real-space positions of atoms comprising the unit cells under the probe. Therefore, if the positions of atoms, and hence the vectors, $\vec{r_j}$, change as a function of position across the sample, the phase factors associated with these atoms also change.

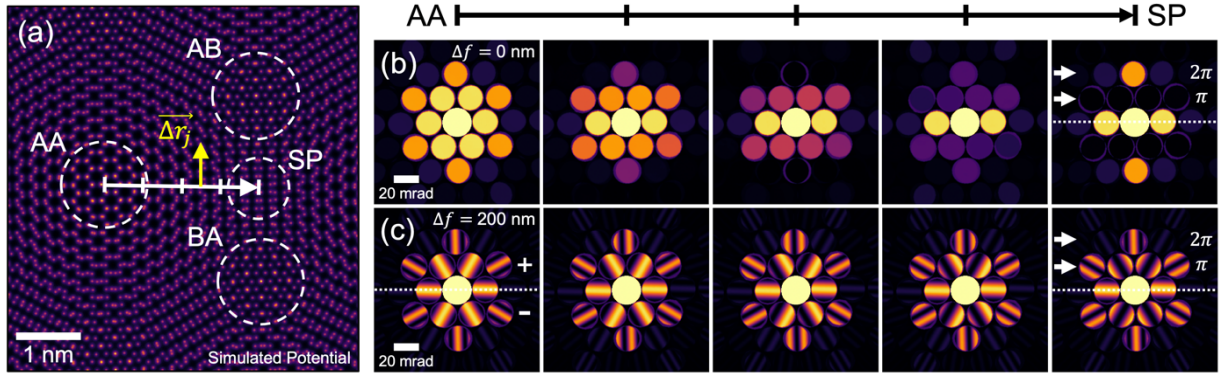

**Figure 2.** Simulated twisted bilayer graphene potential and diffraction patterns traversing an AA to an SP site. (a) Simulated twisted bilayer graphene potential, with a 3° twist angle between the constituent layers. Between AA and SP regions, $\overrightarrow{\Delta r_j}$ increases linearly in the upward direction for a rigidly twisted material. (b) Diffraction patterns from an in-focus probe simulated for steps from AA to SP, approximately indicated by vertical dashes on the line profile in (a). (c) Diffraction patterns from a defocused probe simulated for steps from AA to SP, approximately indicated by dashes on the line profile in (a), showing the movement of interference fringes in the Bragg overlap regions.

By defining a unit cell for twisted bilayer graphene as one of monolayer graphene but with twice the length in the $c$ (out-of-plane) direction to include both layers of the material, this concept can be used to understand the change in phase of the interference fringes corresponding to changes in real-space structure. For example, as the probe moves from an AA site, where the layers directly overlap, toward a neighboring SP site, where one layer is shifted by half a unit cell with respect to the other, the difference in $\vec{r_j}$ vectors of the layers, or $\overrightarrow{\Delta r_j}$, points in the direction of the shifting atoms and grows linearly for a rigidly twisted structure, as simulated in **Figure 2**. As a result, the relative phase between the diffracted beams changes in the direction of $\overrightarrow{\Delta r_j}$, giving a phase shift of $\pi$ to the interference fringes in the first order Bragg disk overlap regions (in those disks where $\overrightarrow{\Delta r_j} \neq 0$) and $2\pi$ to the second order disks. When using a probe focused in the plane of the sample, the nearly flat intensity of the overlap regions transitions between constructive interference and complete destructive interference, modulating the intensity in the overlap region between a maximum and zero. These phase shifts produce the effect shown in Figure 2b, generating a distinct diffraction pattern at each point on the material. Since each type of overlap region produces a unique diffraction pattern, multivariate statistical analysis techniques can be used to map regions primarily associated with AA, AB/BA, and SP regions, as shown in **Figure S1**. This provides valuable information about the underlying structure, but the oscillation of flat intensity in the overlap region alone does not directly reveal the relative phase between the disks, which would allow additional structural information to be precisely and directly extracted from the data.

In contrast to an in-focus probe, a sufficiently defocused probe produces an interference fringe with a wavelength visible in the overlap region of the Bragg disks, as discussed previously. In this case, the relative phase change of the disks is identical to the in-focus case, but here the change manifests as a visible movement of the position of the fringe within the overlap region, as shown in Figure 2c. Directly viewing interference fringes with shorter wavelengths such as this provides information about the sign of the phase change, its wavelength, and its direction, all of which provide valuable information, as described below, and none of which are available in the in-focus case. Changes in position of the interference fringes are directly proportional to shifts in relative position of the atoms in the layers diffracting under the probe, and the ability to see the direction of the phase shift provides the ability to make an absolute measurement of the twist direction if the sign of the defocus is known. In addition, deviations from a linear ramp of the phase shift across the sample directly reveal structural distortions of the material away from rigidly twisted layers.

To acquire data of this type, we perform 4D-STEM to record a 2D array of diffraction patterns. Optimized beam parameters are used to maximize fitting accuracy, including a high beam current for increased signal-to-noise and the largest convergence semi-angle that does not result in overlap of neighboring disks (in this case, ~10 mrad) to provide the largest area possible for fitting of fringes. An example experimental annular dark-field (ADF)-STEM image from a bilayer graphene material with a moderate twist angle of ~3° is shown in **Figure 3a**. For this image, a larger convergence angle was used to improve resolution and the contrast varies across the moiré pattern of periodic overlaps due to the periodic modulation of intensity in the Bragg disks discussed above, revealing the positions of the AA and AB/BA stacking regions. Figure 3b shows diffraction patterns from a 4D-STEM dataset taken in another region using the optimized beam parameters discussed above. The mean diffraction pattern is subtracted here to minimize effects of the central disk, which is also masked, and to provide interference fringes that oscillate around zero rather than a finite value. Figure 3c shows the sum of pairs of opposite Bragg disks at individual probe positions between an AA and SP site, corresponding approximately to the positions simulated in Figure 2. To find the phase of the interference fringes, the Bragg disks are masked just inside of their edges to remove contributions from non-overlapped regions and a cosine function masked in the same way is subsequently fit to the disks. An appropriate amplitude of the fitting function can be set directly from the experimental data. If the defocus and overall twist angle are known, a fixed value of the wavelength can be set as well (with only small deviations from this value resulting from atomic lattice distortions). Alternatively, the wavelength can be fit directly from the data, which allows all of the key fitting parameters to be directly obtained from the 4D dataset. For example, in this case, fitting the interference fringe wavelength, along with the measured average twist angle of ~1.3° (obtained from AA moiré spot center-to-center distances in an image reconstructed from the 4D data) allows an experimentally-determined defocus value of ~320 nm to be calculated using Equation 3 in the Experimental Section. Obtaining the defocus value directly from experimental data in such a manner increases the robustness of the technique to errors since precise instrumental defocus calibrations are therefore not required and any defocus nonlinearities are accounted for. In addition, small rotations of the interference fringes should occur due to position-dependent variations in interlayer spacing arising from out-of-plane structural relaxations. While these, and hence the interlayer spacing, are measurable in some cases currently (see discussion below), the magnitude of these rotations is small enough here that structural distortion mapping results are not significantly affected. To verify this, we manually added an artificial 2-degree rotation to the interference fringe fitting functions, an order of magnitude larger than the rotations caused by variations in interlayer spacing for the structure in Figure 4, and reanalyzed the data in Figure 5. Standard deviations of the differences induced in the results were more than an order of magnitude smaller than the features shown, confirming that rotations produced by varying interlayer spacings currently contribute negligibly to in-plane displacement mapping results.

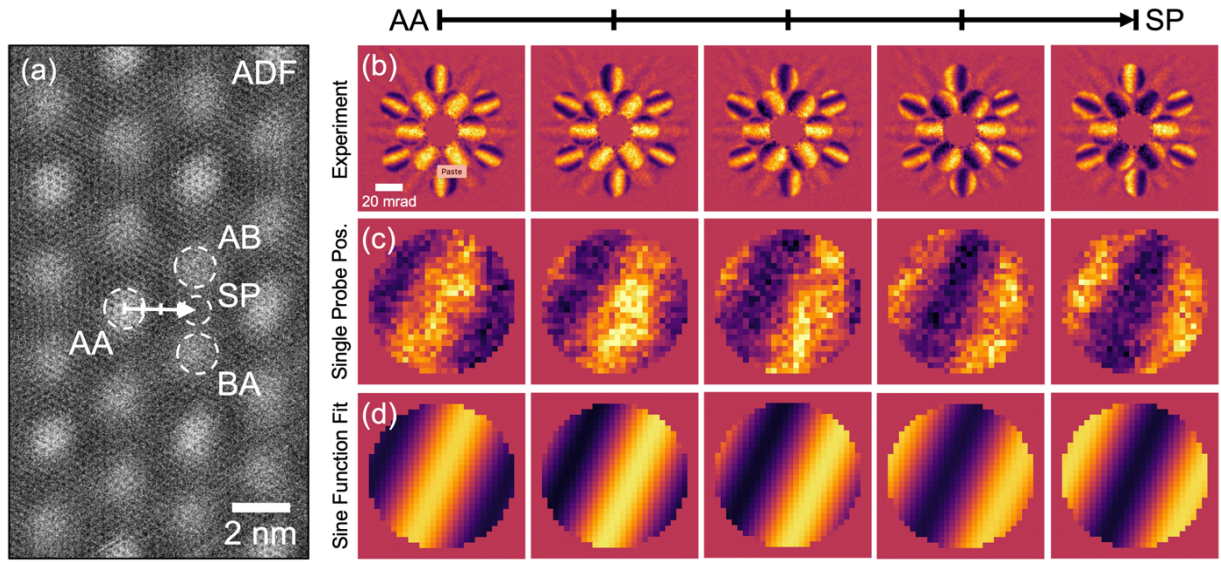

**Figure 3.** Experimental ADF and 4D-STEM data, and fits of the diffraction data, for points between an AA and an SP site. (a) ADF-STEM image of a representative twisted bilayer graphene region with a ~2.8° twist, calculated by comparing moiré and atomic lattice spacings using an FFT. (b) Experimental diffraction patterns from 4D-STEM data taken in another region at an experimentally measured defocus value of ~320 nm, with mean diffraction pattern removed. In this area, the layers producing the diffraction interference are twisted by ~1.3°, as shown in Figure 5. (c) Summed pairs of opposite Bragg disks corresponding to diffraction data in (b), masked slightly smaller than the convergence angle to eliminate effects of non-overlapping regions when fitting. (d) Fits to experimental data in (c) used to generate maps of lattice distortions for full 4D datasets.

Performing the phase fitting procedure for each pair of Bragg disks at every probe position results in three images of the fringe phase periodically ramping from $-\pi$ to $\pi$. Unwrapping the phase of one of these images results in an approximately linear phase ramp due to the overall twist angle of the material, with deviations arising in principle from distortions of the atomic positions away from those that a rigid twist would produce. These images can then be combined using the relationship between the phase change and $\overrightarrow{\Delta r_j}$ with the linear ramp removed, resulting in a map of atomic position distortions. For examples of these intermediate steps, see **Figure S2**.

To explore the capabilities of our technique, we paired van der Waals density functional theory (DFT) calculations[35] with multislice electron diffraction simulations[36–38] to simulate a 4D-STEM dataset for bilayer graphene with a commensurate 3.15° twist angle (see Supporting Information for details on the methodology). The DFT calculations generated the relaxed structure shown in **Figure 4a**, where the black and red arrows display distortions of the top and bottom layer atom positions, respectively, away from a rigidly twisted structure. Interestingly, the results show that the top and bottom layers distort separately on opposite sides of the AA sites, with a maximum displacement of 4.6 pm, and little distortion observed fully surrounding the AB and SP sites. A simulated 4D-STEM dataset was then produced from this structure using similar beam parameters to the experimental data and the magnitude and direction of the lattice distortions measured by the technique outlined above, as shown in Figure 4b,c. The overall features of the relaxed structure are accurately reproduced, with displacements around the AA site and none surrounding the AB or SP sites. The magnitudes of the measured features are approximately half those of the actual structure, however, due to averaging of the distortions under the finite-sized probe. Fluctuations in atom positions resulting from errors in fitting were not observed on the same scale as the lattice distortion features in these maps (picometers), suggesting the fit is very precise for high signal-to-noise data. As shown in **Figure S3**, data with noise included results in random fluctuations of the fits, leaving the overall features unaffected but somewhat obscuring finer details. From these displacements, local deviations in twist angle can be mapped as well, as shown in Figure 4d.

Interestingly, this level of precision enables us to see the initial stages of the structural distortions commonly observed for these materials at lower twist angles. Namely, additional twist around the AA sites, reduced twist around AB sites resulting a triangular expansion of the area, and the beginnings of SP filaments that stretch between neighboring AA sites.

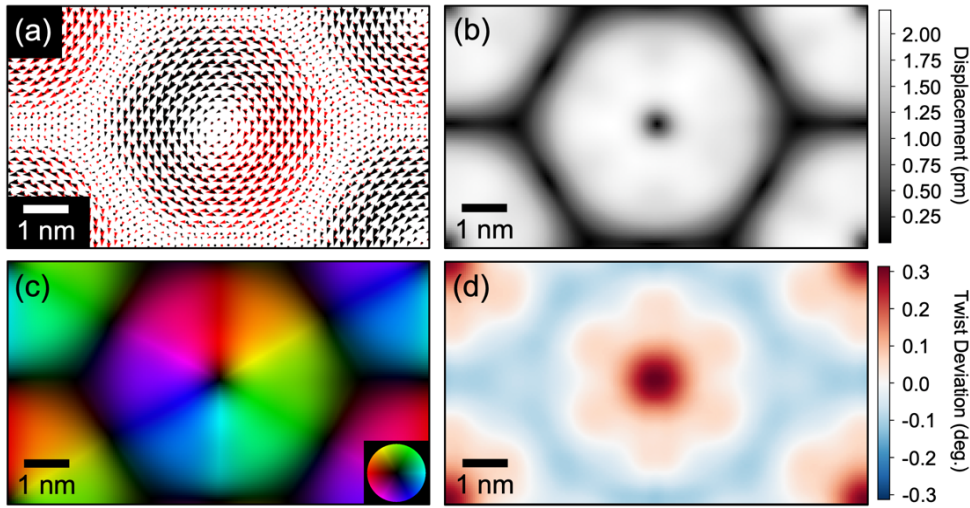

**Figure 4.** Paired DFT structural optimization and multislice electron scattering calculations of twisted bilayer graphene. (a) Relaxed structure produced by DFT, with black and red arrows indicating displacements of atoms in the top and bottom layers, respectively, with the largest arrow indicating 4.6 pm of displacement with respect to positions in a rigidly rotated bilayer. (b) Magnitude of relative displacements observed after multislice simulation of 4D-STEM data from the DFT-provided structure in (a), using the interferometric technique described here. (c) Magnitude (as in (b)) and direction, indicated by the color wheel, of relative displacements of atoms measured by interferometric 4D-STEM performed on simulated data from the structure in (a). (d) Local deviations in twist angle due to distortions in (c).

Finally, as mentioned previously, the interference fringes contain information about the out-of-plane structure as well. Referring again to the schematic in Figure 1, if two layers of a material diffract from different z-heights, $D$ will be different for each layer, and hence the relative direction from one coherent virtual source to the other will be altered from the case of a material with no z-height difference. This results in a rotation of the interference fringes within the Bragg overlap region (and a negligible change in overall wavelength, for small z-height differences). In principle, this rotation can be measured at every probe position to generate an interlayer spacing map. In practice, however, the noise level of current instrumentation precludes this analysis. As detector technology improves, this may indeed become possible. Average information about interlayer spacing may be extracted at present, however, and in some cases provides useful information that is unavailable to other projection techniques. One such case is that of a twisted trilayer graphene material, as shown in **Figure 5**, where information about interlayer spacing allows us to directly determine the stacking sequence of the material. Simultaneously, the ability to precisely measure lattice distortions using the same dataset allows us to confirm the in-plane structural arrangement that results from the stacking sequence, providing structural information in a total of three dimensions.

Figure 5a shows an ADF-STEM image of a trilayer graphene region with relative twist angles of ~1.35° and ~30° between pairs of layers, as confirmed by center-to-center distances of the AA moiré spots in an image reconstructed from the 4D data and directly from the diffraction pattern in Figure 5b, respectively. Three possible stacking sequences exist for a trilayer structure such as this, as outlined in Figure 5c. In two of the cases, the low-angle layers are adjacent to each other in the out-of-plane direction and in the third case they are separated by the high-angle layer. Utilizing the information in the interference fringes and the technique described above, we can definitively discriminate between the two possibilities.

First, an average interlayer spacing can be calculated by finding the average rotation of the interference fringes in the overlap regions. To make this measurement, we averaged diffraction from the three AA sites arranged horizontally in Figure 5a, subtracted the overall mean diffraction data, and averaged over the six symmetric rotations of the resulting pattern.

The result is displayed in Figure 5b. The same fitting procedure as described above was then used to calculate an average fringe rotation of 5.2°. This fringe rotation is directly linked to the interlayer spacing, $d$, through:

$$d = 2\Delta f \sin(\theta_t/2) \tan \theta_f \qquad (2)$$

for small twist angles and fringe rotations, where $\Delta f$ is the defocus, $\theta_t$ is the twist angle, and $\theta_f$ is the fringe rotation angle. The average interlayer spacing resulting from the fringe rotation in Figure 5b is 7.0 Å, around twice that of a typical graphene bilayer. The direction of the fringe rotation additionally allows us to determine the direction of twist of the top layer with respect to the bottom layer, here clockwise when viewed in the beam propagation direction. Since the observed interlayer spacing is around twice that of bilayer graphene, this directly reveals that the two low-angle layers are separated by the third high-angle layer, depicted as situation 3 in Figure 5c. The placement of the high-angle layer between the low-angle layers should hypothetically shield them from interacting with each other, preventing strong periodic lattice distortions.

To confirm that the high-angle layer between the low-angle layers shields them from interacting with each other, we used the same dataset to generate the corresponding relative atom position displacement and local twist angle deviation maps, as shown in Figure 5d and 5e. It is immediately apparent that no distortions with periodicities equal to the underlying lattice are present. Two general features can be observed, however. First, an overall background of increasing displacements and local twist angle is present from left to right in the maps. This indicates a slowly increasing overall twist angle across the field of view, increasing by ~0.13°. Examining the center-to-center distance for AA sites on either side of the ADF-STEM image in Figure 5a confirms that a higher twist angle is present on the right side of the image, corresponding almost exactly to the result from the interferometric technique, albeit with a significantly lower spatial resolution. This establishes the ability for the technique to map variations in twist angle on the order of tenths of a degree, occurring over many nanometers. In addition to this slowly varying overall twist angle, we also observe higher-frequency fluctuations in the maps of displacement amplitude, with a mean of 2.0 pm, and local twist angle deviation, with a standard deviation of 0.06 degrees (after removal of the slowly-varying background). These features appear to be mainly related to a decreased precision of phase fits near integer multiples of a phase of $\pi/4$ (see Figure S4), though some features remain unaccounted for. These features therefore establish the noise level of the real experiment. With this noise level, it should be possible to observe displacements associated with twist angles at least as high as 3°, such as those shown in Figure 4. Since low-angle layers adjacent to each other would result in significantly larger structural distortions than the 3° twist results shown above, this confirms that the high-angle layer is present between the low-angle layers and shields them from each other, resulting in a lack of periodic lattice distortions to within the accuracy of our measurement.

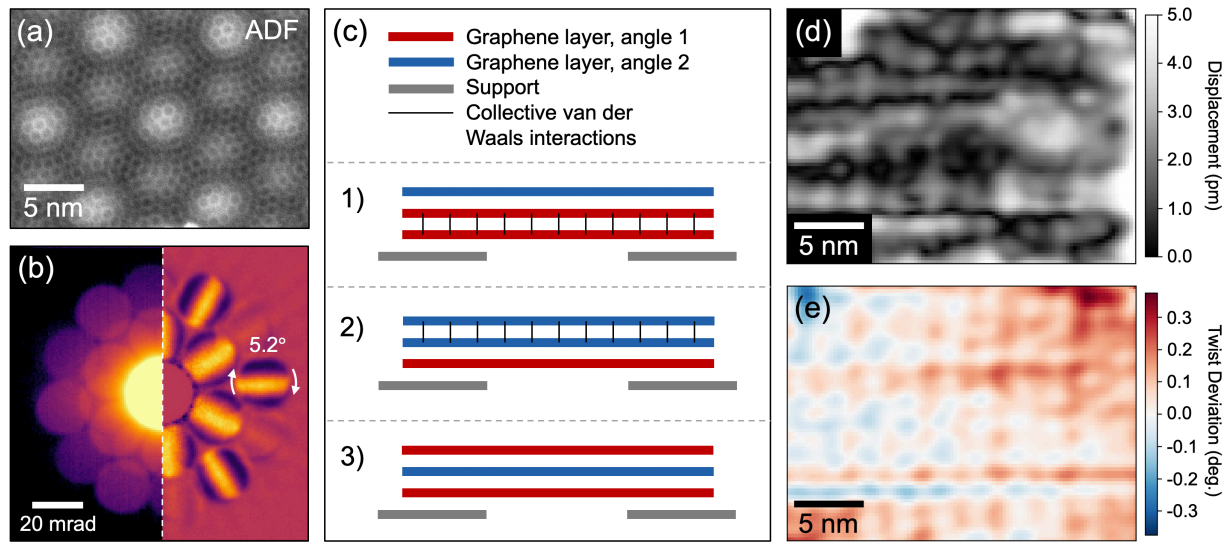

**Figure 5.** Experimental determination of stacking sequence of a twisted trilayer graphene. (a) ADF-STEM image of trilayer graphene with ~1.35° and ~30° twists. (b) Overall average diffraction pattern (log of intensity) and average diffraction taken from the three AA sites situated horizontally in (a), with the overall average subtracted. This latter data is also averaged for each symmetric rotation of the pattern. The average interference fringe rotation is labeled and leads to a measured interlayer spacing of ~7.0 Å. (c) Schematic of possible stacking sequences of a trilayer such as this. The low-angle interlayer spacing measured in (b) reveals that sequence 3 occurs in this case, with the high-angle layer present between the low-angle angle layers. (d) Measurement of atomic displacements from the region in (a), measured by interferometric 4D-STEM. In this case, low-frequency information was removed to more clearly display local fluctuations by filtering the data with a 2D Gaussian function with full width at half maximum of 10 nm and subtracting this from the full data. (e) Local deviation in twist angle due to displacements in (d) without the 10 nm Gaussian filter.

To further confirm that high twist angles do not produce long-range periodic structural distortions and can therefore shield the low-angle layers from each other as observed, we relaxed a commensurate bilayer structure with a 27.8° twist angle.[39] In this case, the small distances and consistent directions to nearest neighbor atoms in separate layers needed to produce collective van der Waals interactions do not occur, and thus long-range structural reconstructions are not present. In addition, the largest deviation of a single bond distance from the mean is ~40 fm, which shows that only very small distortions on individual atoms are present. This confirms that the high-angle layer present between the two low-angle layers can adequately shield the low-angle layers from each other, eliminating long-range periodic structural distortions due to interactions between them while not inducing any itself.

Since no discernable structural distortions should be produced in this case, this afforded us the unique opportunity to establish the noise level of our experiments as well. While our current experimental noise level allows observation of structural distortions in materials with twist angles up to at least 3°, this is by no means an inherent limit of the technique. It is possible that distortions could be measured at higher twist angles, but that would place greater demands on the instrumentation, since the size of distortions decreases with increasing twist angle, requiring more precise measurements to identify. For these experiments, a CMOS indirect electron detector was used to acquire the data. As detector technology continues to improve and direct electron detectors such as the pnCCD,[40] MediPix,[41] and EMPAD[42] become more widely available, measurements with increased sensitivity and decreased noise will become more common, thus potentially allowing lattice distortions down to one pm or less in magnitude to be measured. Ultimately, this may allow the phase, wavelength, and angle of the interference fringes to all be accurately measured across the sample, enabling the relative positions of atoms in separate layers of few-layer materials, such as twistronic materials, to be mapped in all three dimensions with nanometer-scale resolution.

## 3. Conclusions

We have introduced an interferometric 4D-STEM technique that utilizes interference fringes in overlapping Bragg disks to precisely probe relative positions of atoms in different layers of few-layer materials. Through paired DFT calculations and 4D-STEM multislice simulations, we described how this technique is capable of locally mapping the pm-scale lattice distortions that occur in twisted bilayer graphene at a moderate misorientation angle of 3° with nanometer-scale resolution. We additionally demonstrated how this technique allows the twist direction between interfering layers to be obtained, as well as a direct measurement of the average interlayer spacing, which can be used to identify stacking sequences of few-layer materials in certain situations. Further, we established the noise level of our experiments, which is currently limited primarily by the sensitivity and dynamic range of the detector. In principle, performing this interferometric 4D-STEM technique with an optimal detector will allow the relative three-dimensional positions of atoms in few-layer materials to be mapped with nanometer resolution, providing information not only about in-plane lattice distortions but also local fluctuations in out-of-plane interlayer spacing. This method therefore provides a means to better understand the complex interplay between the precise 3D structure of few-layer materials and their electronic state, enabling greater control over tuning of the unique electronic properties that these materials possess.

## 4. Experimental Section

### 4.1 Material Synthesis and STEM Sample Preparation

*4.1.1 Synthesis*. Monolayer graphene was grown on electropolished Cu foil by CVD method. The Cu foil was heated to 1000 °C with $H_2$ (15%)/Ar mixture gas at ~ 1 torr. After a 20 min. anneal, 10 sccm $CH_4$ was introduced for graphene growth for 20 min. After growth, the $CH_4$ gas was switched off and the Cu foil was removed from the heating zone rapidly for fast cooling.

*4.1.2 STEM Sample Preparation.* Poly(methyl methacrylate) (PMMA) was spin-coated on graphene/Cu at a speed of 3000 rpm for 1 min. The Cu foil was then dissolved by iron chloride ($FeCl_3$) solution. After rinsing in DI water several times, the PMMA/graphene was scooped by a Protochips TEM *in situ* heating chip before drying overnight. Finally, the PMMA was removed using acetone and IPA. A second layer graphene was then transferred onto the first by repeating the same steps as described above, to form a few-layered material. This process leaves some amount of residual volatile and non-volatile materials on the TEM sample. *In situ* heating of the sample in the microscope (described below) removes the volatile material, while the non-volatile material remains. Interestingly, although one would assume that even a very small amount of this material would preclude structural measurements from being made on the very thin underlying bilayer material, measurements are still possible through thin regions of this residue. This is due to the fact that the interference fringes originating from the bilayer material are still present in the diffraction patterns and the effect of scattering from the residue is simply an increase in randomly distributed background intensity. For a thin enough material, the interference fringes can still be observed on this background and fit. This does decrease the precision of measurements, however, so clean regions must be used for high-quality results. Nevertheless, this allows us to observe that generally the residue does not significantly affect the structure of the bilayer material (e.g., see Figure S2, top). As a result, we observe that the main effect of the residue is to limit the area usable for generating the highest-quality results.

## 4.2 Density Functional Theory Calculations

For generating a twisted bilayer graphene model structure, we first created a commensurate rigidly rotated supercell with a misorientation angle of 3.15 degrees between the layers (containing 662 carbon atoms) and applied 25 Å of vacuum between the periodic images in the out-of-plane direction. Similar to the methodology of Lucignano and colleagues,[37] we then employed density functional theory to relax the structure with a van der Waals functional to obtain an *ab initio* prediction of the in-plane and out-of-plane reconstructions of the two layers. We used the real-space projector-augmented wave simulation package GPAW,[43,44] and tested both the DF2[45] and C09[46] vdW functionals (the former overestimates interlayer spacing, while the latter overestimates the binding energy[47]) before settling on the latter. High computational efficiency was achieved with the help of the libvdwxc library[48]. We used a real-space computational grid spacing of 0.2 Å, a 3×3×1 Monkhorst-Pack **k**-point mesh for Brillouin-zone integrations and enforced a strict maximum force convergence criterion of 0.01 eV/Å to ensure precisely relaxed geometries. As an additional high-misorientation angle bilayer structure, we created a commensurate 27.8° twist angle unit cell with 52 atoms following Ref. [39] and relaxed it using the same methodology.

### 4.3 4D-STEM Simulations

The relaxed structures obtained from the DFT calculations were used as input for image simulations using the multislice algorithm[49] implemented in the abTEM code.[38] The potential was calculated with the independent atom model using finite projection integrals and the parametrization of Lobato et al.[50] The structure was repeated to obtain a Fourier space sampling of 0.0125 Å$^{-1}$. The real space sampling was 0.05 Å, and the slice thickness was 0.1 Å. The scan area covered the unique positions of the structure with a step size of 0.2 Å. The convergence angle was 8 mrad. Neither partial coherence nor thermal diffuse scattering was included in the simulations.

### 4.4 STEM Experiments

STEM experiments were performed on a Nion UltraSTEM 100, operated at 60 kV. The sample was placed on a Protochips *in situ* heating chip and heated to ~1200 °C for two hours in the STEM to remove volatile contamination. The sample was then returned to room temperature before data was acquired. In some cases, *in situ* heating such as this has been shown to induce local changes in the stacking order of bilayer graphene,[22] while in other cases it did not.[51] In our experiments, we did not observe changes in the material structure due to extended heating, possibly due in part to residual material from the sample transfer "pinning" the bilayer in place. High-resolution ADF data was recorded using a beam convergence semi-angle of ~31 mrad and a medium-angle annular dark-field detector with minimum collection angle set just outside of the central disk. For 4D-STEM measurements, a convergence angle of ~10 mrad was used to maximize Bragg disk areas without causing overlap between neighboring disks from a single layer of the material. A high current mode was used to maximize the signal-to-noise ratio of the recorded disks, with a probe current of ~120 pA. 4D-STEM diffraction data was recorded on a Hamamatsu ORCA CMOS detector. The total detector size was 2048x2048 pixels, and data was acquired using the central 320x320 pixels, binned by a factor of two. At smaller detector resolutions such as this, the detector allows acquisition times down to ~1 ms, but 10 ms was chosen for this work to maximize signal without causing saturation. 128x128 real-space probe positions were used, for a total acquisition time of approximately three minutes.

## 4.5 Data Processing

Data was processed using standard packages in Python, such as Numpy, Scipy, etc. As stated in the main text, each opposing pair of Bragg disks was averaged and a circular mask applied just within the outer edge of the disks, in this case an 8 mrad radius, setting everything outside of this radius to zero to avoid information from the non-overlapped portion of the disks. In principle, the interference fringes contain all of the information to find the relative positions of the atoms in separate layers in three dimensions through the phase, wavelength, and angle of the fringes, but in this section we focus on revealing in-plane distortions through the fringe phase. To find this phase, we set the interference fringe wavelength in reciprocal space using known values of defocus and overall twist angle of the material (found through an ADF, image, for example) with the following, which can be obtained by examining the geometry of the scattering in Figure 1:

$$\theta_{\lambda,int} = \frac{\lambda_e}{2\,\theta_B\,\Delta f\,\sin(\theta_t/2)} \tag{3}$$

for $\lambda_e \ll 2\,\theta_B\,\Delta f\,\sin(\theta_t/2)$ and neglecting the often small contribution due to the interlayer spacing, where $\lambda_e$ is the wavelength of the beam electrons, $\theta_B$ is the Bragg scattering angle, $\Delta f$ is the defocus, and $\theta_t$ is the twist angle. Small deviations from the overall twist angle due to structural distortions have little effect on the fitting of the phase of the fringes. Alternatively, the optimal wavelength can be fit from the data. In addition, the amplitude of the fringes can be found through "virtual dark-field" maps from the Bragg disk pairs. Finally, due to the noise level of current instrumentation, the direction of the fringe is set as constant and as with the twist angle, small deviations due, for example, to fluctuations in interlayer spacing have little effect on the fitting of the fringe phase. The average rotation angle can be found as discussed in the main text, however. Given the average interference fringe wavelength and rotation, a cosine function can then be fit to each of the three pairs of fringes at every probe position, identifying their phase. The resulting periodic phase fits, as shown below in Figure S2c, are then unwrapped to form nearly linear ramps, which are combined and converted to real-space relative positions through the relationship in Equation 1 above. Knowing the relative positions of the atoms in each layer also allows local deviations in twist angle to be calculated. Partial derivatives of the relative displacement vectors can be used to form a matrix in the same form as a strain tensor, which allows the local twist angle to calculated in the same manner that lattice rotation would be with a strain tensor,[52] through the difference of cross terms (change of x displacements in the y direction, and vice versa). The resulting mean value gives the overall twist of the material, and deviations from this value give local variations in the twist angle, as shown in Figure 4d, S3e, and 5e.

**Supporting Information**

Supporting Information is available from the Wiley Online Library or from the author. abTEM codes used to generate simulated 4D-STEM data are available at: https://github.com/jacobjma/abTEM/tree/master/articles/2021_small_interferometric_4D-STEM

**Acknowledgements**

This research was supported by the Center for Nanophase Materials Sciences, which is a U.S. Department of Energy (DOE) Office of Science User Facility. M.C. is supported by DOE-BES early career award ERKCZ55. A.M. and T.S. were supported by the European Research Council (ERC) under the European Union's Horizon 2020 research and innovation programme (Grant agreement No. 756277-ATMEN), and acknowledge computational resources provided by the Vienna Scientific Cluster. We would like to thank Andrew R. Lupini and Juan Carlos Idrobo for very helpful discussions.

# Supporting Information

**Title:** Interferometric 4D-STEM for Lattice Distortion and Stacking Sequence Measurements of Few-layer Two-dimensional Materials

*Michael J. Zachman\*, Jacob Madsen, Xiang Zhang, Pulickel M. Ajayan, Toma Susi, Miaofang Chi\**

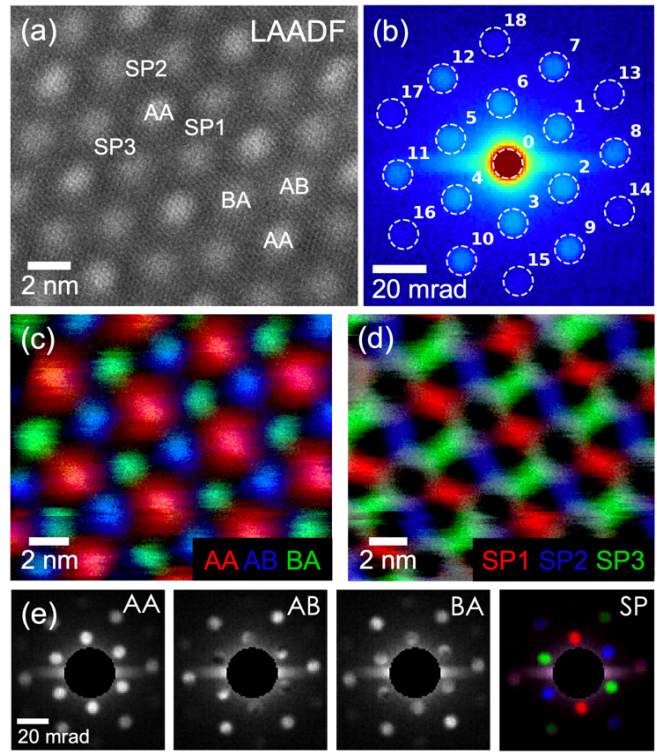

**Figure S1.** Mapping of regions with distinct real-space structure by multivariate curve resolution (MCR) analysis of reciprocal-space information in a 4D-STEM dataset. (a) Low-angle annular dark-field (LAADF) image of twisted bilayer graphene with a twist angle of ~3°, with overlap regions of distinct symmetry labeled. (b) Average diffraction pattern acquired from the region shown in (a), using a semiconvergence angle of ~4 mrad. (c) Real-space components corresponding to AA, AB, and BA overlap regions resulting from MCR analysis of the full 2D array of diffraction patterns acquired in the region of (a). (d) Real-space components for the three SP orientations resulting from MCR analysis. (e) Corresponding reciprocal-space components for the AA, AB, BA, and three SP overlap regions (with SP colors paired with to the real-space component colors in (d)).

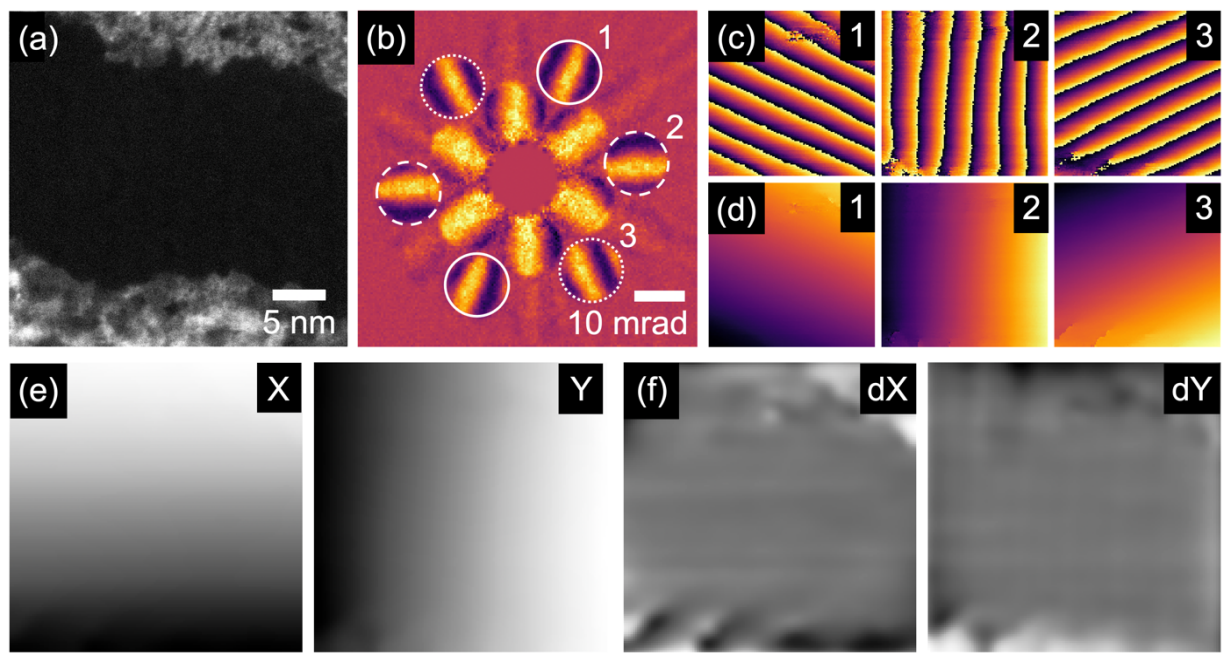

**Figure S2.** Intermediate steps in calculating distortions in relative lattice positions. (a) High-angle annular dark-field image of the region used for the results show in Figure 5, acquired in-focus. (b) Example defocused diffraction pattern acquired near an AA region, with opposite pairs of Bragg disks to be used for fitting labeled. (c) Phase results from fitting the pairs of Bragg disks labeled in (b). Note that these maps show that measurements can be made through thin regions of the residue remaining from the transfer process, though the precision is sufficiently reduced that high-quality results cannot be obtained here. Nevertheless, these maps show that the residue does not significantly alter the underlying bilayer structure. (d) Results from (c) after phase unwrapping. (e) Results from (d) after conversion to relative real-space positions of the interfering layers in both the X and Y directions. (f) Results from (e) after removing planes ramping linearly in the Y and X directions, leaving only deviations from this linear ramp (labeled dX and dY).

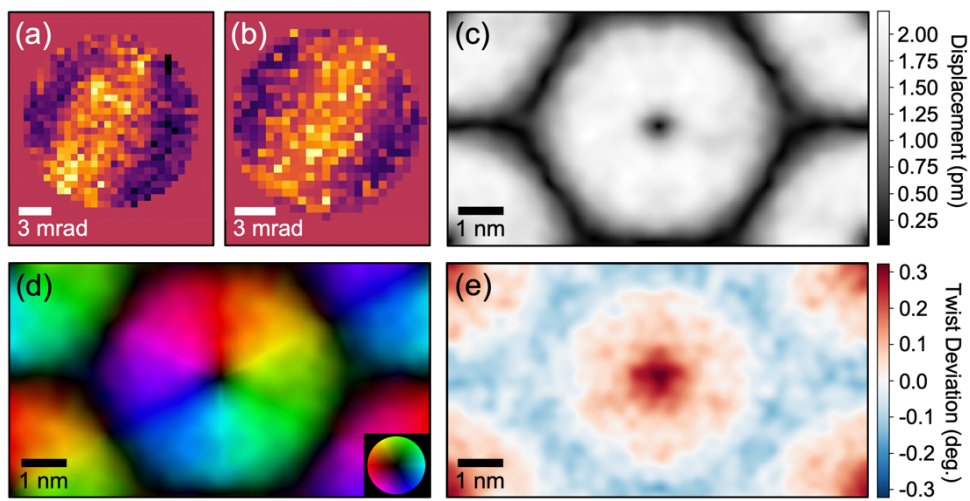

**Figure S3.** Effect of noise on measurements. (a) Example of a single experimental Bragg disk. (b) Example of a single simulated Bragg disk with noise comparable to the experimental data included. (c) Magnitude of relative displacements measured by phase fitting method, applied to the simulated data in Figure 4 with noise added as in (b). A ~1 Å real-space Gaussian filter was applied post-analysis for visualization. (d) Corresponding magnitude (as in (c)) and direction of relative displacements. (e) Local deviations in twist angle due to distortions in (d).

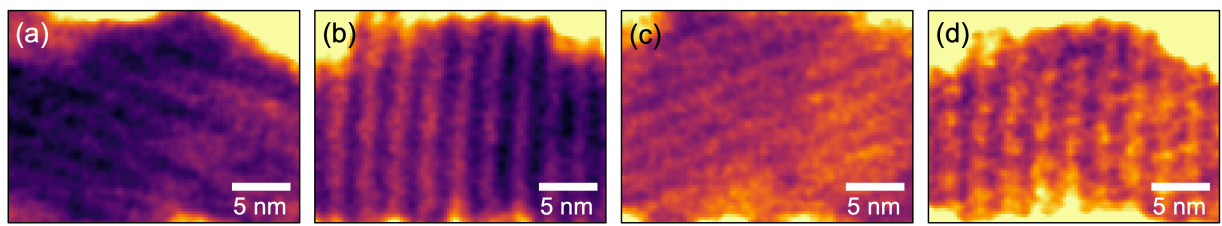

**Figure S4**. Standard deviations of phase fits for the analysis displayed in Figure 5. (a)-(c) Standard deviations for each of the three pairs of opposing Bragg disks. (d) Combined standard deviations from (a)-(c). Note the decreased precision near the edges due to proximity to residual contamination from transfer. Maps in (a)-(c) are displayed from 0.3 pm to 0.6 pm and in (d) from 0.9 pm to 1.2 pm.